# *ON THE ORIENTATION OF THE HISTORIC CHURCHES OF LANZAROTE: WHEN HUMAN NECESSITY DOMINATES OVER CANONICAL PRESCRIPTIONS*

Alejandro Gangui,[1] A. César González García,[2]
Mª Antonia Perera Betancort,[3] and Juan Antonio Belmonte[4]

(1) Instituto de Astronomía y Física del Espacio (CONICET-UBA), Buenos Aires, Argentina.
(2) Instituto de Ciencias del Patrimonio, Incipit, CSIC, Santiago de Compostela, España.
(3) Servicio de Patrimonio Histórico. Excmo. Cabildo Insular de Lanzarote, Canarias, España.
(4) Instituto de Astrofísica de Canarias, La Laguna, Tenerife, España.

Abstract: The orientation of Christian churches is a well-known distinctive feature of their architecture. There is a general tendency to align their apses within the solar range of rising directions on the horizon, favoring orientations close to the east (astronomical equinox), although the alignments in the opposite direction, namely, with the apse towards the west, are not unusual. The case of the churches built in northwest Africa before the arrival of Islam is representative in this regard, and may reflect earlier traditions. As the Canary Islands are the western end of this North African common culture, it is relevant to study compact sets of churches located in the different islands of the archipelago. We started our project with Lanzarote, seeking for pre-European traditions –including astronomical ones– being merged with new ones as in other aspects of Canarian culture. We have measured the orientation of 30 churches built prior to 1810, as well as a few buildings of later times, nearly a complete sample of all the island's Christian sanctuaries. The analysis here indicates that a definite orientation pattern was followed on the island but, unlike what is often found in most of the Christian world, it has two interpretations. On the one hand, the representative orientation to the east (or west) is present. However, the sample has also a marked orientation towards north-northeast which is, as far as we know, a pattern exclusive to Lanzarote. We analyze the reasons for this pattern and suggest that one possible explanation could be a rather prosaic one, namely, that sometimes needs of everyday life are more relevant than –and push individuals to make decisions at odds with– religious beliefs. This work is the beginning of the first systematic archaeoastronomical study ever conducted with old churches in the Canary Islands.

## Introduction

The study of the orientation of Christian churches has been of interest for long time and has recently received new impetus in the literature as it was recognized that it represents a key feature of their architecture.[1] According to the texts of the early Christian writers and apologists, churches' apses should lie along a particular direction, that is, the priest had to stand facing eastward during services. This is recognized by Origen, Clement of Alexandria and Tertullian, and the first Council of Nicaea (AD 325) decreed that it should be that way. Also in the fourth century, St. Athanasius of Alexandria declared that the priest and participants should face east, where Christ, the Sun of Justice, would shine at the end of time (*'ecclesiarum situs plerumque talis erat, ut fideles facie altare versa orientem solem, symbolum Christi qui est sol iustitia et lux mundi interentur'* [...]). A thorough analysis of the early sources and methods of orientation can be found in the Bibliography.[2]

However, these requirements are ambiguous, making it possible to choose between different interpretations: should the church be oriented towards sunrise on the precise day its foundations were prepared? Or at the sunrise of another day that may have been relevant, such as the feast day of the saint to whom the church was originally dedicated? Or was the orientation to be strictly towards the east? Were the churches oriented to sunrise at the equinox? In this case, toward which equinox? There are several possibilities:[3] the Roman vernal equinox occurred on March 25, while the Greek equinox was on March 21, as was reflected in the Council of Nicaea; but other definitions might also have been in use, such as the entrance of the Sun in the sign of Aries (which might be different from the canonical equinox as established by the church) or the autumnal equinox. From each of these definitions we would obtain dates, and therefore orientations, that are slightly different.[4]

---

[1] González-García, A.C. and J.A. Belmonte, 'The orientation of pre-Romanesque churches in the Iberian Peninsula', *Nexus Network Journal*, Vol. 17 (2015), pp. 353-377.

[2] Vogel, C., 'Sol aequinoctialis. Problèmes et technique de l'orientation dans le culte chretien', *Revue Sciences Religieuses*, Vol. 36 (1962), pp. 175-211; McCluskey, S.C., *Astronomies and cultures in early Medieval Europe* (Cambridge: Cambridge University Press, 1998).

North Africa, in spite of the Roman dominance, is an exception to the rule. In many regions of Africa, such as Proconsularis and Tripolitania, a number of churches with orientations towards the west -which is a usual custom in the early times of Christianity- is found.[5] These regions are relevant for our study, because they are possible homelands of the Canarian aboriginal population. Note also that most of these churches are oriented within the solar range (with orientations between the winter and summer solstices), with clustering around the equinoxes and solstices.

We started a large-scale project in the Iberian Peninsula and the Canary Islands. In the latter area, this work is the first such systematic study. Our interest is to check the orientation of the churches of Lanzarote, as it can provide us with a compact set of old churches where we can search for pre-European or canonical religious traditions, including astronomical ones, or a mix of both. This could provide a broader understanding of one key aspect of Canarian culture.[6]

Churches and chapels of Lanzarote—a description

Religious architecture on the island of Lanzarote began with the building of modest chapels endowed with a single room. In some, over time, small shrines or altars were added in their headers, together with vestries on their sides and other elements of practical use, such as low walls bordering the atrium (barbicans) and calvaries (Figure 1). In general, these constructions were not subjected to strict building rules; thus, their structure was erected according to the needs of the moment. Nowadays, small chapels are in general located far from the cities, while some of the churches eventually achieved some monumental dimensions, like those in Teguise and Arrecife.

Given the large number of historical monuments, close to thirty and therefore suitable for a statistical analysis, Lanzarote was chosen as the first test area to study the orientation of the Canarian churches in the immediate post-conquest centuries (XV century onwards). The research aim was to analyze whether in this location the orientation of the monuments was influenced by factors such as the presence of aboriginal people in the islands, who had worship habits and calendric systems completely different from the newly arrived people.[7]

---

[3] McCluskey, S.C., 'Astronomy, Time, and Churches in the Early Middle Ages', in M.-T. Zenner (ed.), *Villard's legacy: Studies in Medieval Technology, Science and Art in Memory of Jean Gimpel* (Aldershot: Ashgate, 2004), pp. 197-210.

[4] Ruggles, C.L.N., 'Whose equinox?', *Archaeoastronomy*, Vol. 22 (1999), pp. S45-50; González-García, A.C. and J.A. Belmonte, 'Which Equinox?', *Archaeoastronomy, The Journal of Astronomy in Culture*, Vol. 20 (2006), pp. 97-107.

[5] Esteban C., J.A. Belmonte, M.A. Perera Betancort, R. Marrero and J.J. Jiménez González, 'Orientations of pre-Islamic temples in North-West Africa', *Archaeoastronomy*, Vol. 26 (2001), pp. S65-84; Belmonte J.A., A. Tejera, M.A. Perera and R. Marrero, 'On the orientation of pre-Islamic temples of North-west Africa: a reappraisal. New data in Africa Proconsularis', *Mediterranean Archaeology and Archaeometry*, Vol. 6, no. 3 (2007), pp. 77-85; Belmonte J.A., M.A. Perera Betancort and A.C. González-García, 'Análisis estadístico y estudio genético de la escritura líbico-bereber de Canarias y el norte de África', in *VII Congreso de patrimonio histórico: inscripciones rupestres y poblamiento del Archipiélago Canario* (Arrecife: Cabildo de Lanzarote, 2010), in press.

[6] Belmonte J.A. and M. Sanz de Lara, *El Cielo de los Magos* (Islas Canarias: La Marea, 2001).

[7] Belmonte, J.A., C. Estéban, A. Aparicio, A., Tejera Gaspar and O. Gónzalez, 'Canarian Astronomy before the conquest: the pre-hispanic calendar', *Rev. Acad. Can. Ciencias*, Vol. 6, no. 2-3-4 (1994), pp. 133-156.

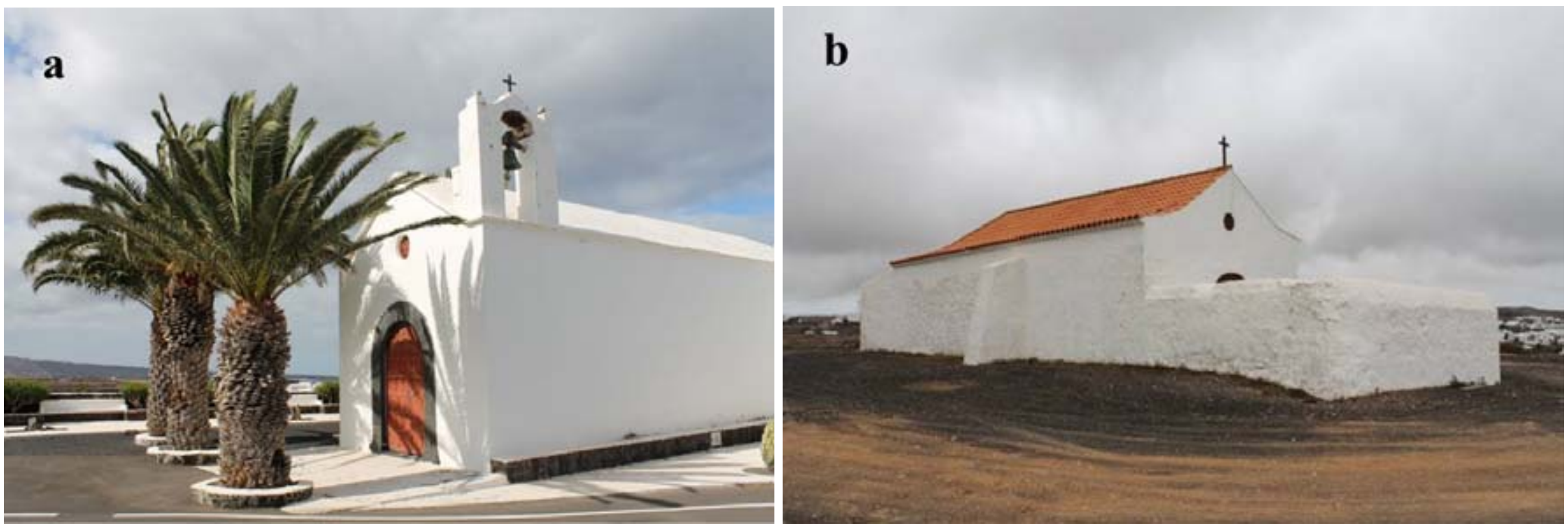

*Fig. 1: Two churches in Lanzarote with unique features: (a) the church of Nuestra Señora de las Mercedes (Our Lady of Mercy) in Mala is the only one on the island with a precise equinoctial orientation (nearly 90° of azimuth), and that may also be oriented to the solar sunrise on the day of its Marian devotion (September 24), a not very frequent feature in the island environment; (b) the chapel of San Rafael, located alone and isolated on a plateau overlooking El Jable on the outskirts of the village of Teguise; it dates from the XIX century but it is already cited in 1661. Its orientation is eastward (72°.5) and has the peculiarity that it possesses a large "L"-shaped barbican that protects the entrance from the prevailing winds, as we can infer from the sand that is seen deposited on it.*

Table 1 shows the data obtained in our campaign of fieldwork. The identification of the churches and their coordinates are presented along with their orientation (archaeoastronomical data): the measured azimuth (rounded to 1/2° approximation) and the angular height of the point of the horizon towards which the apse of the church is facing, as well as the corresponding computed declination. Regarding the construction dates provided for the churches, there is some ambiguity: in general, the year is only estimated and it corresponds to the first reported reference of the building (and for one church it has not been clarified). This dating can be useful for our later analysis of the data.

| LOCATION | NAME / DATE | L (°/') NORTH | l (°/') WEST | a (°) | h (°) | δ (°) | PATRON SAINT DATE / Orientation |
|---|---|---|---|---|---|---|---|
| (1) Femés | San Marcial de Limoges (1630) | 28/54 | 13/46 | 52½ | 1¼ | 32¾ | 16 Abril - 8 Julio / ---- |
| (2) Yaiza | Ntra. Sra. de los Remedios (1699) | 28/57 | 13/46 | 77½ | 2¾ | 12 | vv.ff. 9 Sept / 21 Abr – 22 Ago |
| (3) Uga | San Isidro (1956) | 28/57 | 13/44 | 15 | 3½ | 60 | 15 Mayo / ----- |
| (4) Masdache | La Magdalena (s. XX) | 28/59 | 13/39 | 339½ | -0½ | 54¼ | 22 Julio/ ----- |
| (5) La Geria | Ntra. Sra. de la Caridad (1706) | 28/58 | 13/43 | 44½ | 0½ | 38¾ | 8 Sept R 15 Agos / ----- |
| (6) Mancha Blanca | Ntra. Sra. de los Dolores (1782) | 29/02 | 13/41 | 92 | 1¾ | -1¼ | 15 Sept / 17 Mar – 26 Sept |
| (7) Tinajo | San Roque (1669) | 29/04 | 13/40 | 34½ | -0¾ | 45¼ | 16 Agos R/ ----- |
| (8) Yuco | Ntra. Sra. de Regla (1663) | 29/03 | 13/39 | 356½ | -1¼ | 59 | vv.ff. / ----- |
| (9) Tiagua | Ntra. Sra. Del Socorro (1625) | 29/03 | 13/38 | 12 | -1¼ | 57¼ | 8 Sept / ----- |
| (10) Sóo | San Juan Evangelista (1749) | 29/05 | 13/37 | 142 | 0¾ | -43½ | 27 Dic / ----- |
| (11) Tao | San Andrés (1627) | 29/02 | 13/37 | 6½ | -1 | 58¾ | 30 Nov / ----- |
| (12) Mozaga | Ntra. Sra. de la Peña (1785) | 29/01 | 13/36 | 278 | 1¼ | 7½ | R 8-13 Agos / 8 Abr – 4 Sept |
| (13) San Bartolomé | San Bartolomé (1661) | 29/00 | 13/36 | 26½ | 0 B | 51 | 24 Agos R / ----- |
| (14) Nazaret | Ntra. Sra. de Nazaret (1648) | 29/02 | 13/34 | 105 | 1¾ | -12¼ | R 26 Agos / 17 Feb – 25 Oct |
| (15) Teguise | San Rafael (1661) | 29/04 | 13/34 | 72½ | 1¾ | 15¾ | 29 Sept / 4 May – 9 Ago |
| (16) Teguise | El Cristo de la Vera Cruz (1625) | 29/04 | 13/33 | 82 | 0 B | 6¾ | 3 Mayo / 7 Abr – 5 Sept |
| (17) Teguise | San Juan de Dios y San Fco. de Paula (Sto. Domingo) (1698) | 29/03 | 13/34 | 254½ | 0½ | -13¾ | 8 Mar – 2 Abr / 11 Feb – 1 Nov |
| (18) Teguise | Ntra. Sra. de Guadalupe (1680) | 29/04 | 13/34 | 128½ | 5½ | -30¼ | 6 Sept / ----- |
| (19) Teguise | Ntra. Sra. Miraflores, Convento de San Francisco (1588) | 29/03 | 13/33 | 84 | 6½ | 8 | ? – 4 Oct / 10 Abr – 2 Sept |
| (20) Teseguite | San Leandro (1674) | 29/03 | 13/32 | 71 | 0¾ | 16½ | 13 Nov / 6 May – 6 Ago |
| (21) El Mojón | San Sebastián (1661) | 29/04 | 13/31 | 42½ | 0¾ | 40¼ | 20 Ene / ----- |
| (22) Los Valles | Santa Catalina (1749) | 29/05 | 13/31 | 339½ | 14¼ | 65¼ | 25 Nov – 20 Abr / ---- |
| (23) La Montaña (Teguise) | Ntra. Sra. de las Nieves (1661) | 29/06 | 13/32 | 15½ | -1 | 56 | 5 Agos / ----- |
| (24) Haría | San Juan (1625) | 29/09 | 13/30 | 98½ | -0½ | -7¾ | 24 Jun / 1 Mar – 12 Oct |
| (25) Haría* | Ntra. Sra. de Encarnación (1631) | 29/09 | 13/30 | 74 | 4½ | 15¾ | 25 Mar / 4 May – 9 Ago |
| (26) Mala | Ntra. Sra. de las Mercedes (1809) | 29/06 | 13/28 | 89½ | -0½ | 0 | 24 Sept / 20 Mar – 23 Sep Equinoccio |
| (27) Guatiza | El Cristo de las Aguas (1915) | 29/04 | 13/29 | 107½ | 1 | -15 | R 13 Sept / 8 Feb – 3 Nov |
| (28) Arrecife | San Ginés (1570) | 28/57 | 13/33 | 53½ | 0 B | 31¼ | R 25 Agos / ---- |
| (29) Tahiche | Santiago Apóstol (1779) | 29/01 | 13/33 | 70 | 12½ | 23 | 25 Jul / 11 Jun – 2 Jul |
| (30) Tías | Ntra. Sra. de la Candelaria (1795) | 28/58 | 13/39 | 323½ | 8¾ | 49¾ | 2 Feb / ----- |
| (31) Conil | María Magdalena (1794) | 28/59 | 13/40 | 118½ | 12 | -18½ | 22 Jul / 27 Ene – 15 Nov |
| (32) Tegoyo | Sagrado Corazón de Jesús (1863) | 28/58 | 13/41 | 52½ | 9¼ | 36¾ | Móvil, Junio / ---- |

*Table 1: Orientations for the chapels and churches of Lanzarote. For each building, we show the location, identification (name and most likely date of construction), the geographical latitude and longitude (L and l), the astronomical azimuth (a) taken along the axis of the building towards the apse (rounded to 1/2° approximation), the horizon angular height (h) in that direction (0 B means horizon is blocked; we take h = 0°) and the corresponding resultant declination (δ). Some of the churches were surrounded by nearby hills; this justifies the high value of angular height (h) for them.*

We obtained our measurements using a pair of tandem instruments which incorporate a clinometer and a compass with a precision of half a degree, and also by analyzing the landscape setting of each of the buildings (see Figure 2). We then corrected the data according to the local magnetic declination (geomag.nrcan.gc.ca/calc/mdcal-en.php). Our values of the magnetic variation for different sites on the island range from 4°38' to 4°46' W. As the precision of the measured magnetic azimuths is about 1/2°, we have used the same magnetic declination (4°42' W) for all 32 values of orientation (this may be considered a limitation in precision but it simplifies the calculation). The obtained values are the average of two and sometimes even three measurements, and we must emphasize that, with few exceptions, the various measurements differed by less than 1/2°. In any case, given the magnetic disturbances recorded at various locations on the island (especially in the vicinity of the volcanic eruptions of the Timanfaya), some measures (one fifth of the total) have been verified with photo satellite images, and we found few differences. Therefore, we estimate the error of our measurements to be around ±3/4° (upper bound), and thus the data is suitable for a statistical study of the monuments' orientation.

In Figure 2(a), we show the orientation diagram for the churches and chapels. The diagonal lines on the graph indicate, in the eastern quadrant, the extreme values of the corresponding azimuth for the sun (azimuths of 62°.5 and 116°.7 -continuous lines-, equivalent to the northern hemisphere summer and winter solstices, respectively) and for the moon (azimuths of 56°.4 and 123°.8 -dotted lines-, equivalent to the position of the major lunistices).

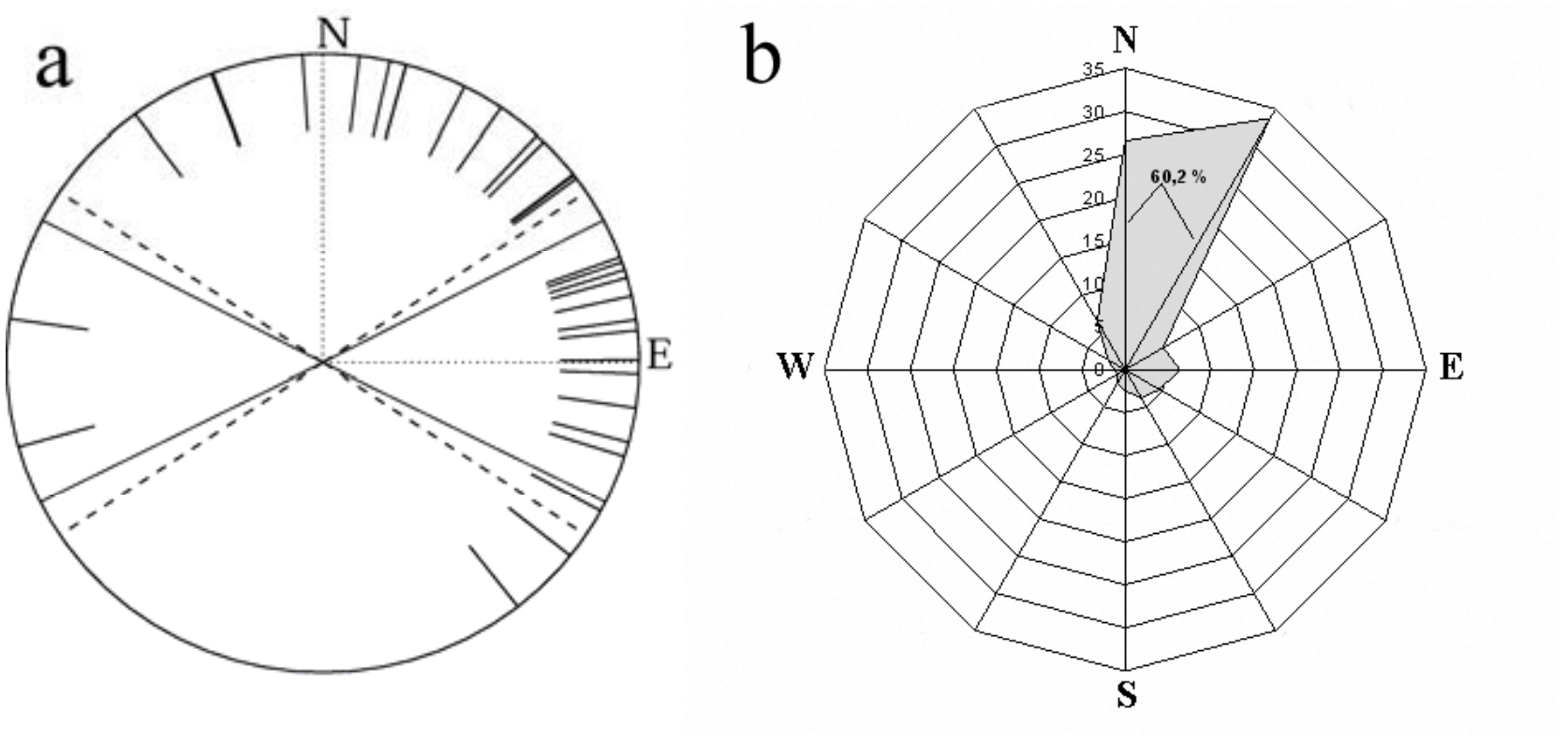

*Fig. 2: (a) Orientation diagram for the churches and chapels of Lanzarote, obtained from the data in Table 1. Although a significant number of monuments follows the canonical orientation pattern in the solar range, a non-negligible number of churches are oriented to the north-northeast. (b) Diagram of winds for Arrecife Airport in Lanzarote, illustrative of the prevailing winds on the island. Note the enormous concentration in the range N-NE (azimuths between 345° and 45°), similar to the exceptional orientations of several churches on the island.*

Of the 32 chapels and churches we measured, 12 are oriented in the northern quadrant (between 315° and 45°), 2 in the western quadrant, 17 are oriented in the eastern quadrant (with 13 of them in the solar range) and only 1 in the southern quadrant; see Figure 2(a). The sample is representative of the island of Lanzarote (although it is not of all the Canary Archipelago), and here two distinct orientations are distinguished: (i) to the North, with "entrance" on the leeward/downwind side, avoiding perhaps the dominant winds of the place, and (ii) eastward, with the apse of the chapel pointing toward the eastern quadrant. The 13 monuments facing *ad orientem* fall within the logic observed in other studies of orientations of churches, but what is remarkable here is the large number of monuments oriented to the northern quadrant, falling outside of the solar range. It seems to be a case singular of Lanzarote where practical and prosaic issues (the orientation against the trade winds from the NNE, see Figure 2(b)) appear side by side with cultic and canonical traditions (i.e., the orientations within the solar range).

While there might be different underlying causal factors for this church orientation pattern, the idea that they may be astronomically oriented is suggestive. Regarding the solar range, there are two particularities. On the one hand we have the mother church of the historic capital city of the island, Teguise, oriented with an azimuth 128°.5, roughly five degrees from the direction (123°.8) of the southernmost rising point of the moon (an orientation already found, in the aboriginal world, in some paradigmatic cases like in the Roque Bentayga, in the island of Gran Canaria, or in the Mount Tindaya, in the island of Fuerteventura).[8] Secondly, the church of Nuestra Señora de las Mercedes (Our Lady of Mercy) in Mala (Fig. 1), is the only one on the island precisely oriented towards the equinox, which also seems to follow a uncommon rule on the island, namely to be oriented to the sunrise on the date of its devotion, as we see in Table 1. (The church Nuestra Señora de los Dolores -Our Lady of Sorrows-, in Mancha Blanca, also seems to be equinoctial, but with more discrepancy.) In this sense, it is also to be noted that most of the churches of Teguise -where most of the initial settlers established- are oriented (with the notable exception of the mother church of Guadalupe, with its "anomalous" orientation, as stated above) to declinations included in the canonical range. While the equinox was also important in the aboriginal world,[9] it seems that in the site which the majority of the European population selected for settlement the rules/habits from their places of origin were respected.

To better understand the above discussion, in Figure 3 we present the declination histogram, which is independent of the geographical location and the local topography. This shows the astronomical declination versus the normalized relative frequency, which enables a clear and more accurate determination of the structure of the peaks. Again, the peak associated with the orientations to the north-northeast, absolutely outstanding, dominates the chart.

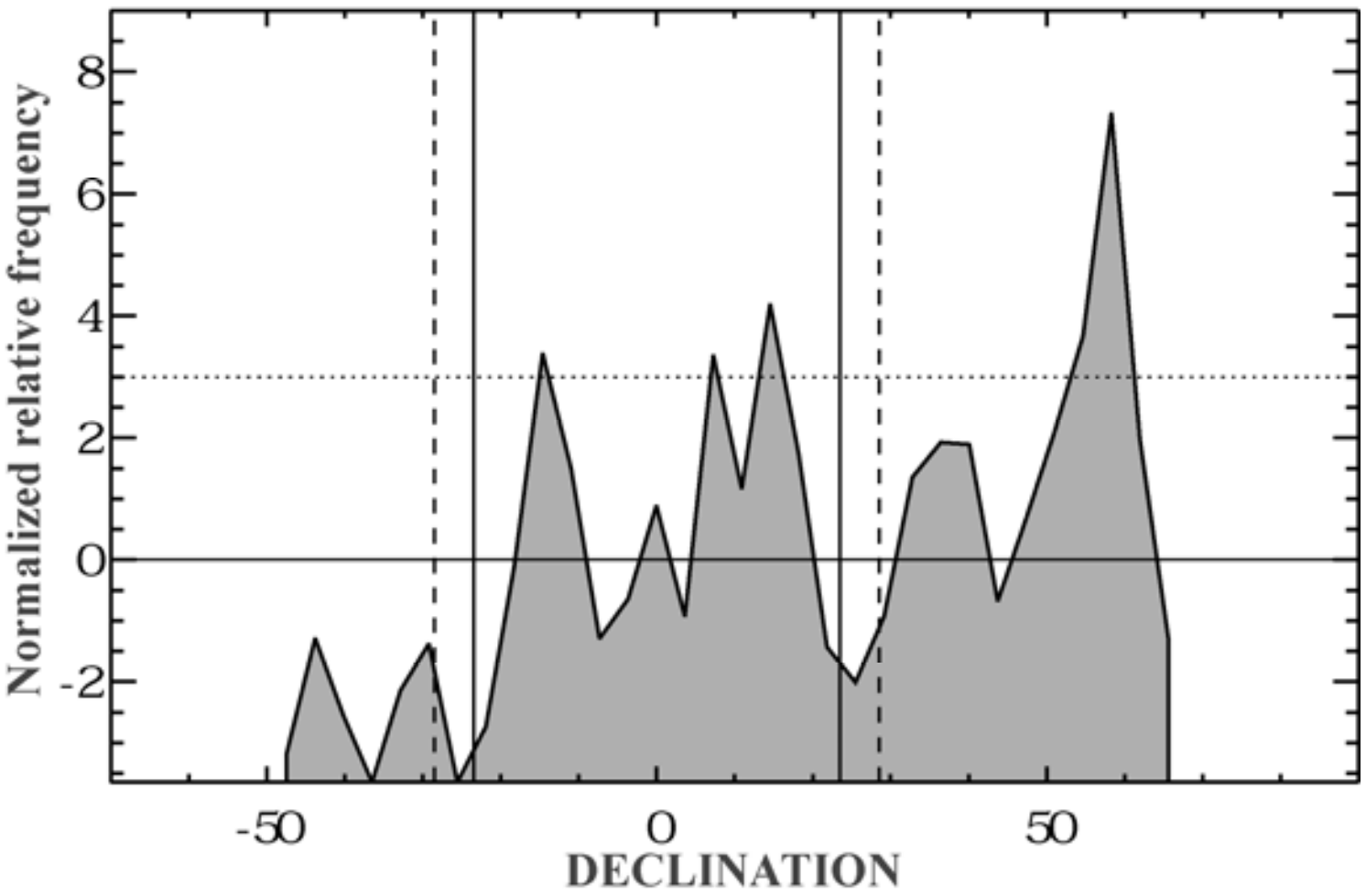

*Fig. 3: Declination histogram for the chapels and churches of Lanzarote. Only a few statistically significant peaks are found above the 3σ level (dotted horizontal line). Continuous vertical lines represent declinations corresponding to the extreme positions of the sun at the solstices, while the dashed vertical lines represent the same for the moon in major lunistices. Three statistically significant minor peaks are found in the solar range (canonical orientation). However, the highest peak, located around 58° is exceptional and is certainly associated with a accumulation peak due to orientations near the meridian line.*

When we encountered this phenomenon in the data, we pondered over the possible causes, establishing different interpretations: could it be the result of the influence of previous aboriginal traditions?[10] Was this related to the settlement on the island of Moorish slaves imported from the nearby African coast and of Islamic tradition?[11] However, it turned out to be that the most prosaic explanation was the one that hinted at the most plausible solution, as we will see in the next section.[12]

---

[8] Belmonte, J.A. and M. Hoskin, *Reflejo del Cosmos: atlas de arqueoastronomía en el Mediterráneo occidental* (Madrid: Equipo Sirius, 2002).

[9] Belmonte J.A., C. Estéban, R. Schlueter, M.A. Perera Betancort and O. González, 'Marcadores equinocciales en la prehistoria de Canarias', *Noticias del IAC*, 4-1995 (1995), pp. 8-12.

[10] Belmonte and Hoskin, *Reflejo del Cosmos*, 2002.

## The orientation of churches and landscape

The particular features of the churches we measured in Lanzarote have little correlation with other studies already mentioned in the previous sections. In the Iberian Peninsula and overall in the Mediterranean, the orientation ranges are predominantly within the solar limits.[13] In particular, the important proportion of churches oriented roughly northward we found is new: examples of this are the church Nuestra Señora de las Nieves (Our Lady of the Snows) in La Montaña, patron saint of the island, and also many others, very old ones, such as those located at Tiagua or Tao. It is noteworthy that a significant proportion of churches oriented in this way belong to the northwestern and central sectors of the island, as shown in Figure 4.

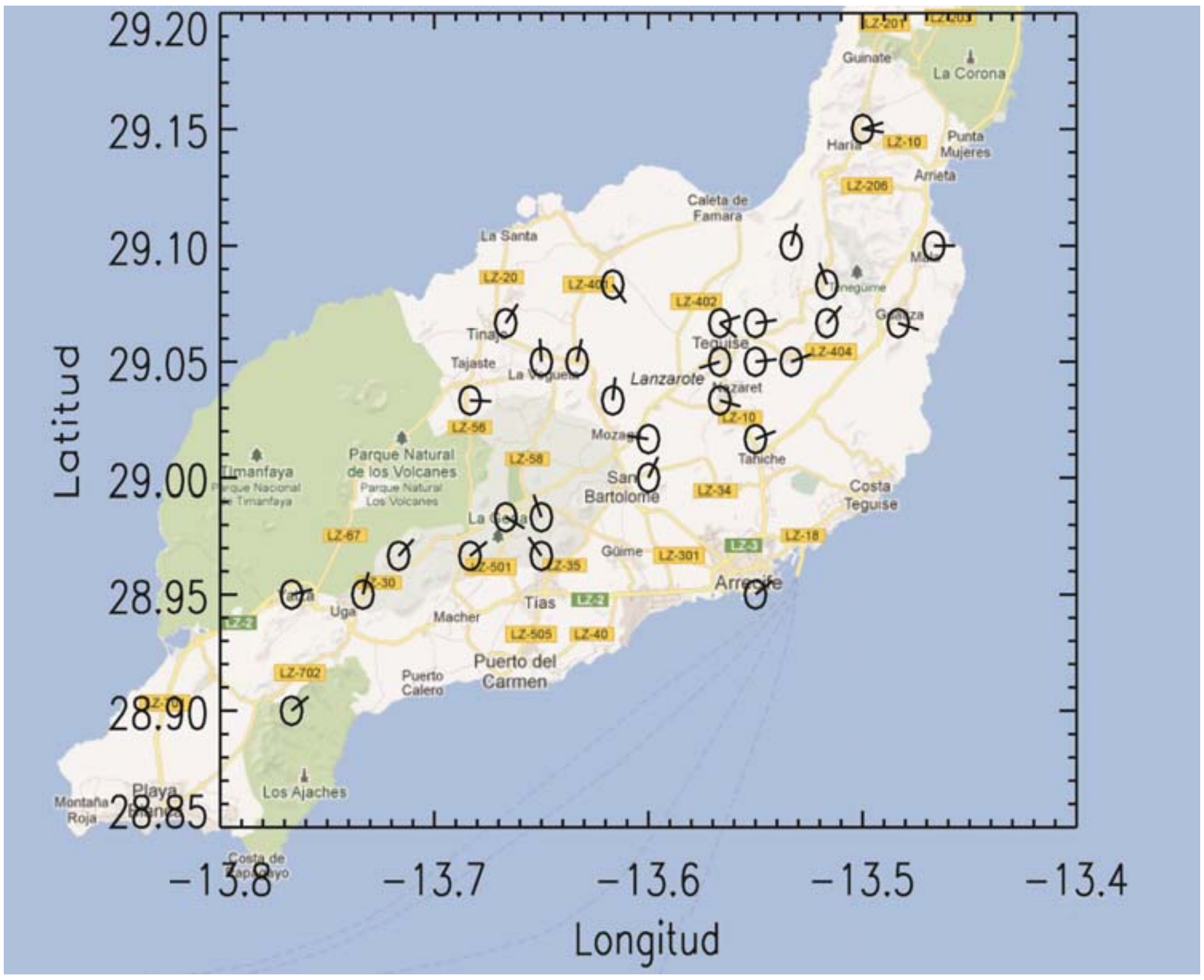

*Fig. 4: Map showing the geographical location of all the measured churches (indicated by ellipses), together with the orientation of the axis of the buildings towards the apse (oriented according to the azimuths recorded in the Table 1). In the town of Haría, two geographically very close churches have different orientations, which explains the presence there of a single ellipse with two stripes. The same happens with a couple of churches in the town of Teguise. Image based on a map courtesy of Google Maps.*

---

[11] Anaya Hernández, L.A., 'El Corso Magrebí y Canarias. El último ataque berberisco a las islas: la incursión a Lanzarote de 1749', in *Actas X Jornadas de Estudios sobre Lanzarote y Fuerteventura* (Santa Cruz de Tenerife: Servicio de Publicaciones del Excmo. Cabildo Insular de Lanzarote, Litografía Romero, 2004), Vol. 1, pp. 13–29; Anaya Hernández, L.A., 'La liberación de cautivos de Lanzarote y Fuerteventura por las Órdenes Redentoras', in *Actas XII Jornadas de Estudios sobre Lanzarote y Fuerteventura* (Santa Cruz de Tenerife: Servicio de Publicaciones del Excmo. Cabildo Insular de Lanzarote, Litografía Romero, 2008), Vol. 1, no. 1, pp. 65–93.

[12] León Hernández, J. de, M.A. Robayna Fernández and M.A. Perera Betancort, 'Aspectos arqueológicos y etnográficos de la comarca del Jable', in *Actas II Jornadas de Historia de Lanzarote y Fuerteventura* (Santa Cruz de Tenerife: Servicio de Publicaciones del Excmo. Cabildo Insular de Lanzarote, 1990), Vol. 2, pp. 283–319.

[13] González-García and Belmonte, 'The orientation of pre-Romanesque churches in the Iberian Peninsula', 2015.

The difference between our results for Lanzarote and those of other studies that have been completed elsewhere,[14] leads us to look for alternatives in trying to understand the pattern of orientations of the churches of this island. If these monuments, in general, are not oriented according to the sun, could it be due to such prosaic reasons as the need to orient the porch of the constructions contrary to the dominant winds coming from the NE direction onto the island (as in Fig. 1) or otherwise protect it? Or else, could it be due to the topography (perhaps changing over time) of different regions of the island? In any case, it seems clear by looking at the graphics and images that the environmental issue is relevant.

Regarding the winds, the areas where more churches facing north-northeast have been built (with their entrance oriented towards the southern quadrant) is on the verge of El Jable (north and center of the island), where it becomes imperative to avoid the sand driven by the wind, sometimes in raging storms as that of 1824 that buried several villages, and that even today, despite the changes in the landscape, shows its lasting effects.[15] Interestingly, the highest number of canonical orientations (i.e., eastward) is found in buildings located in the northeast of the island, in the lee of the wind, in areas protected of the sand by the cliffs of Famara.

In relation to the changing topography of the island, the volcanic eruptions of the Timanfaya that occurred between 1730 and 1736, perhaps also may have played a role. For example, the chapel of St. Catherine in Los Valles, shows a construction of the XVIII century with a peculiar orientation, which replaced a previous one dedicated to the same cult, which had been destroyed during the Timanfaya lavas. The current chapel is oriented with azimuth 339°.5, i.e. in the NNW direction, but is protected by the mountains that surround it, so it may be considered anomalous from any point of view (it would be interesting to know the orientation of the early church). In any case, these ideas seem attractive, as they suggest that human needs can sometimes override obligations of the cult.

To check any noticeable evolution in the characteristics of the buildings over time, we conducted a study on the evolution of the orientation with respect to building age. In Figure 5 we show the values of azimuth and declination versus the probable dates of construction for the churches: 5(a) includes all of the oldest churches for which we know these dates (28 of the 32 in the sample); 5(b) selects those churches oriented within the solar range, i.e. whose azimuths are located between the two horizontal lines (azimuths 62°.5 and 116°.7).

The third panel (c) shows the declination values as a function of date (with horizontal lines at declinations -23°.5 and +23°.5). We note that building orientations stay in the canonical range throughout the entire period; but starting in the second quarter of the seventeenth century, large-scale construction of north-facing chapels began, perhaps in a time when human needs and associated environmental factors far exceeded those of worship. The chart also reflects eruptions on Timanfaya, since the construction of chapels almost ceases in the first half of the eighteenth century, reactivating soon after, but again showing the characteristic pattern of two representative orientations.

[14] Čaval, S., 'Church orientations in Slovenia', in Ruggles, C.L.N., ed., *Handbook of Archaeoastronomy and Ethnoastronomy* (New York: Springer, 2014), pp. 1719-1726.

[15] Perera Betancort, M.A., 'Aportación al problema de El Jable a principios del siglo XIX', in *Actas X Jornadas de Estudios sobre Lanzarote y Fuerteventura* (Santa Cruz de Tenerife: Servicio de Publicaciones del Excmo. Cabildo Insular de Lanzarote, Litografía Romero, 2004), Vol. 1, pp. 205–212.

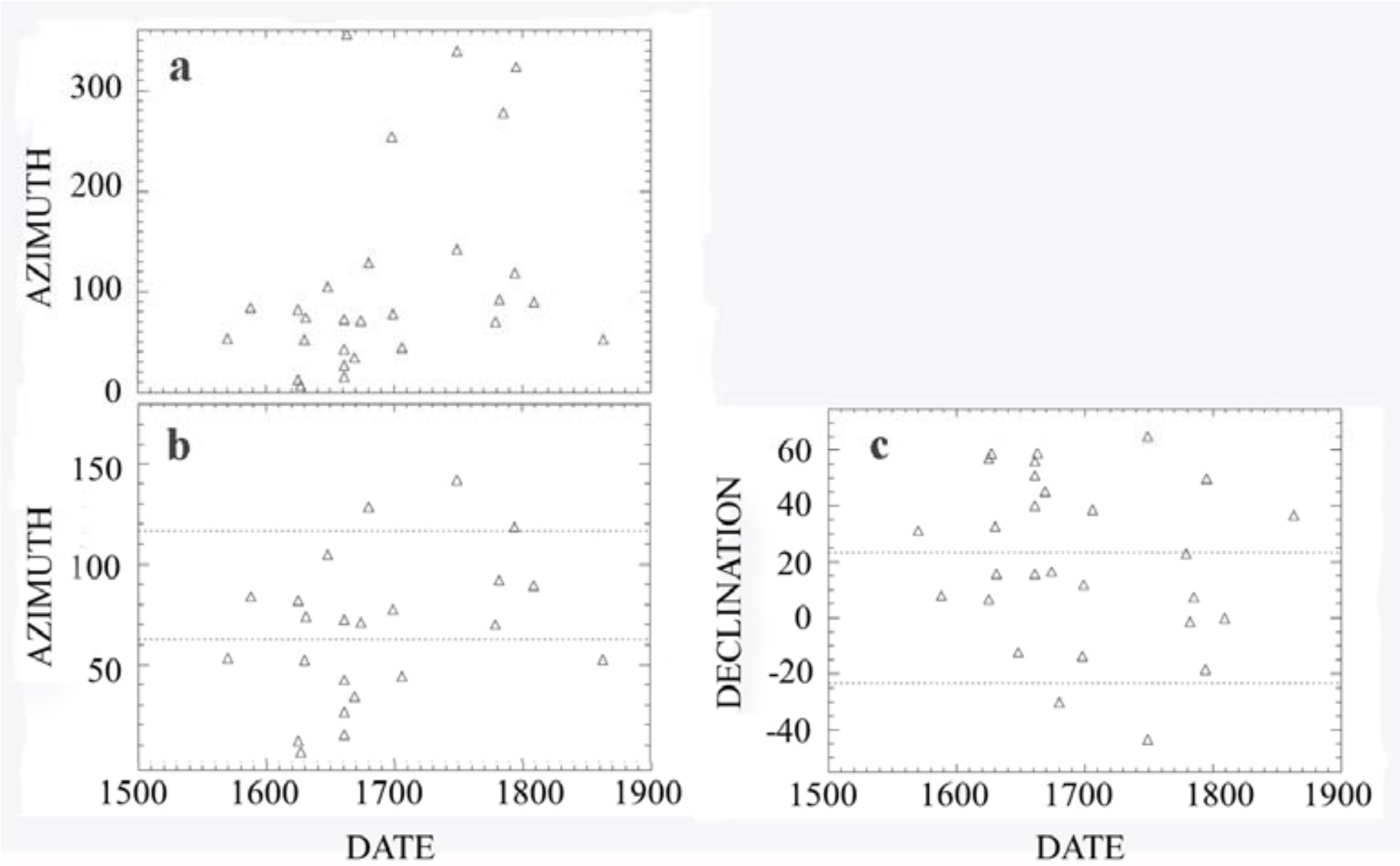

*Fig. 5: (a) Diagram showing the azimuth of the churches and chapels versus the most probable date of construction (or its first mention in the sources). We include 28 out of the 32 buildings studied. (b) Same as the above diagram, but with extended area close to the solar azimuth range. Except for those buildings with a clear orientation towards the N or NE, a good proportion of churches fall within the solar range, as indicated by the two parallel lines. (c) Declination diagram versus the date. The horizontal lines indicate the extreme declinations where the sun is located at solstices. Note the gap in the construction of chapels possibly associated with eruptions of Timanfaya in the first half of the eighteenth century.*

Conclusions

After the conquest and colonization of Lanzarote by the European population in the early fifteenth century, in the decades immediately following it began the large-scale colonization of the island with the establishment of small farms and villages, together with some larger towns like Teguise or Femés. This was accompanied by the construction of a non-negligible number of Christian churches reflecting the new social and religious situation.

It is possible that in a few places, the orientation of sacred buildings followed the pattern of the aboriginal cult, which was centered in the celestial sphere, notably on the sun and the moon. In others, the canonical tradition of aligning the temples eastward was respected (with some exceptions to the western quadrant), although with a much greater degree of tolerance than was usual. In this regard, only one church in Lanzarote, the one in Mala, appears to have an orientation consistent with the solar sunrise on the day of the (Marian) dedication (Fig. 1).

Finally, we find a relatively large number of north-northeast oriented churches in Lanzarote, which is a notable exception to the rule. We analyzed different possibilities to explain this anomaly, and concluded that this pattern of orientation appears to reflect the desire to avoid the strong winds prevailing on the island, which come precisely from that direction, and in particular to minimize the discomfort caused by the sand displaced by the wind on buildings located close or bordering with El Jable.

This is the first stage of a project which will be developed in the coming years, namely measuring the orientation of the oldest Christian churches and chapels in other islands of the Canary Archipelago. For example, it would be interesting to study the orientation of old churches in the island of Fuerteventura, subjected to the same wind flow -even more intense- than that blowing on the neighboring island of Lanzarote. Will the churches of Fuerteventura also show the same pattern of orientations? Have their builders dared to violate the canonical precept to impose the human needs over the cult? Future work will help us to answer this question.